\documentclass[aps,prb,showpacs,superscriptaddress,twocolumn]{revtex4-1}
\usepackage{subfigure}
\usepackage{float}
\usepackage{bm}
\usepackage{amsmath,amsfonts,amssymb,graphicx}
\begin{document}

\title{Multifrequency Excitation and Detection Scheme in
Apertureless Scattering Near Field Scanning Optical
Microscopy}
\author{H. Greener}
\author{M. Mrejen}
\author{U. Arieli}
\author{H. Suchowski}
\email{haimsu@post.tau.ac.il}
\affiliation{School of Physics and Astronomy, Tel Aviv University, Tel Aviv 69978}ֿ
\affiliation{Center for Light-Matter Interaction, Tel Aviv University, Tel Aviv 69978}

\begin{abstract}
We theoretically and experimentally demonstrate a multifrequency excitation and detection scheme in apertureless near field optical microscopy, that exceeds current state of the art sensitivity and background suppression. By exciting the AFM tip at its two first flexural modes, and demodulating the detected signal at the harmonics of their sum, we extract a near field signal with a twofold improved sensitivity and deep sub-wavelength resolution, reaching $\lambda/230$. Furthermore, the method offers rich control over experimental degrees of freedom, expanding the parameter space for achieving complete optical background suppression. This approach breaks the ground for non-interferometric complete phase and amplitude retrieval of the near field signal, and is suitable for any multimodal excitation and higher harmonic demodulation.
\end{abstract}


\maketitle
\section{Introduction}
In the past few decades, tremendous progress has been made in optical imaging beyond the diffraction limit \cite{Novotny, ALewis, Pohl}. Near field microscopy has revolutionized this field, as it allows for noninvasive and nondestructive retrieval of deep sub-wavelength optical information, providing unprecedented information on optical properties of materials at the nanoscale \cite{BETZIG1468, Inouye, Zenhausern, KnollNFProbing, historynsom, lipson}. Thus the field has opened a window to phenomenon such as fundamental light-matter interactions, chemical reactions and transport phenomenon in two-dimentional materials \cite{phononpolaritons, guybartal, basovhBN, basovgraphene, RaschkeChemImaging, eladgross, vdvabajo}. The apertureless version of the scattering near field scanning optical microscope (sSNOM) has expanded to the optical regime the topographic probing capabilities of the atomic force microscope (AFM). The sSNOM utilizes  the AFM's sharp tip by dithering its amplitude in the proximity of a sample and illuminating it by focused light \cite{knoll2000enhanced, NSOMHillenbrand, KeilmannHillenbrand2004}. Owing to the nonlinearity of the light scattering process with respect to the tip-sample distance, high harmonic demodulation allows near field imaging with a spatial resolution mainly limited by the apex of the tip \cite{knoll2000enhanced}. However, to date, a thoroughly background-free image necessitates implementing various schemes, such as pseudoheterodyne detection, an interferometric technique in which a phase modulated reference enables the extraction of the pure near field signal\cite{ocelic2006}. 

Recently, it has been shown \cite{PhysRevLett.100.076102, AFMbook} that mechanically exciting the AFM cantilever at two or more of its first flexural modes results in enhanced force sensitivity and improved resolution to topographic images \cite{PhysRevLett.99.085501}. The coupling between the two mechanical modes is at the origin of this so-called multifrequency AFM sub-atomic spatial resolution, as the higher harmonics of the first mode acts as an effective driving force for its higher eigenmodes \cite{HigherHarmonicsGenAFM}.

Here we implement the multifrequency AFM scheme to increase the near field optical sensitivity of the sSNOM. Our newly formulated theoretical model is based on the bi-modal excitation of the AFM tip as described in Figure \ref{fig:Figure1}(a) and a multifrequency detection scheme in SNOM (MF-SNOM). It predicts a set of experimental parameters relevant for the suppression of optical background in the detected signal. We observed that in the multimodal excitation method, the solution space for these parameters spreads over a two-dimensional plane, thus allowing further degrees of freedom in near-field measurements. We experimentally show that this scheme allows for a further enhanced sensitivity in the measurement of a near field signal as a function of tip-sample distance. Since this z-axis sensitivity is directly related to the in-plane spatial resolution in such measurements \cite{zresolution}, this technique could potentially lead to enhanced resolution in the x-y plane. We believe that this is a feasible method that will allow for enhanced sensitivity, improved resolution and background-free near field images.

\section{Theoretical Model}
\begin{figure}
\includegraphics[scale=0.12]{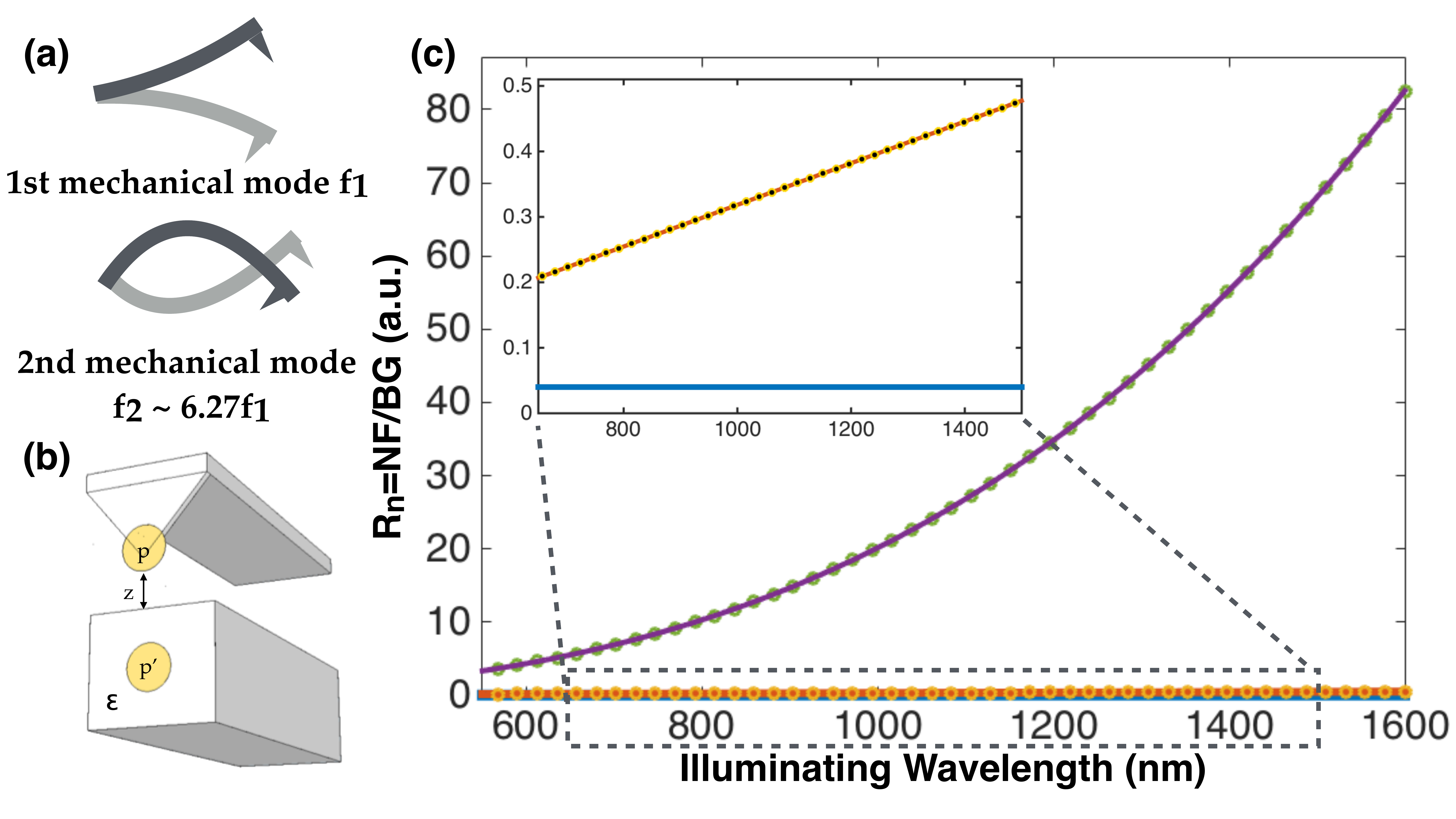}
\protect\caption{(Color online) (a) Simultaneous bi-modal excitation of cantilever in multifrequency near field scanning optical microscopy at two first flexural frequencies, where commonly $f_{2}\approx 6.27f_{1}$ \cite{freecantilever}. (b) Tip-sample system modelled as a polarized sphere of radius $a$, oscillating at amplitude $A$ at small distances $z$ from sample of dielectric constant $\epsilon$. (c) Near-field to background optical contrast enhancement factor $R_{n}$ as a function of illuminating wavelength $\lambda$ for different demodulation frequencies $n$. $R_{n}$ is enhanced for higher demodulation frequencies. The enhancement increases as a function of $\lambda$. Purple: $R_{n}$ calculated for mono-modal excitation and signal demodulation at $4\omega$. Green circles: $R_{n}$ calculated for bi-modal excitation and signal demodulation at $2(\omega + \omega')$. Inset: $R_{n}$ enhancement for low monomodal demodulation frequencies exhibits a slower increase. Blue: $R_{n}$ calculated for mono-modal excitation and signal demodulation at $\omega$. Red: $R_{n}$ calculated for mono-modal excitation and signal demodulation at $2\omega$. Yellow circles: $R_{n}$ calculated for bi-modal excitation and signal demodulation at $\omega+\omega'$. \label{fig:Figure1}}
\end{figure}

The theoretical basis of our model employs a quasi-electrostatic approach for the tip-sample system. The tip is modeled as a sphere of radius $r$, which is imaged in a sample of dielectric constant $\epsilon$,  set at a distance of $z$ away from the tip, as shown in Figure \ref{fig:Figure1}(b). Using the method of images , one can calculate an effective polarizability $\alpha_{eff}$, and apply Mie scattering theory, to calculate the electromagnetic field scattering
cross-section of the probe tip $C_{scatt.}=\frac{k^{4}}{6\pi}|\alpha_{eff}|^{2}$, assuming its radius is smaller than
the illuminating wavelength \cite{knoll2000enhanced}. This scattering cross-section, which
is the weak near field signal of interest, is a non-linear function of the distance
between the probe and sample, by virtue of $\alpha_{eff}$. Varying the tip-sample distance with time leads to a significant modulation of the above near-field scattering coefficient from the tip, while the scattered light from the cantilever body remains constant \cite{ashnichols}. Due to this reason, by demodulating
the detected scattered signal at the higher harmonic frequencies of the cantilever's
motion, one could achieve a narrower cross-section, with a more abrupt
change of signal, as the tip approaches the sample. This is equivalent to effectively sharpening the
probe tip. Nevertheless, while this process results in higher near
field sensitivity to optical measurements, there is a trade-off, since
the measured signal becomes significantly weaker.

In order to gain a more intuitive understanding of the above, we employed a simplified scattering model \cite{labardi2000artifact} to express the bi-modal
near field scattering amplitude in an intuitive form: $K[z(t)]  =  exp\{-z(t)/d\}$.

In our work, the bi-modal motion of the tip at its first two flexural frequencies $f_{1}=\omega/2\pi$ and $f_{2}=\omega'/2\pi$  is represented by $z(t)=Acos(\omega t)+Bcos(\omega' t)$, and $d$ is the typical distance for which the near field term decays.
The $z$ motion artifact due to optical interference (background) is $W[z(t)] = sin[\frac{4\pi z}{\lambda}+\frac{\pi}{4}]$.
Thus, the detected signal function is a sum of the above $S(t)=W(t)+bK(t)$,
where $b$ is the scattering weight, dependent on the scaling of the scatterer volume, which in our case,
is the spherical tip.

One could assume that the tip excitation amplitudes and $d$ are
much smaller than the illuminating wavelength and expand the signal
to order $O(x^{4})$. Separating the different frequency terms results
in a series of Fourier coefficients detected with a Lock-In amplifier,
such as:

\begin{eqnarray*}
S(t) & \approx & DC\\
 & + & \left[\frac{1}{\sqrt{2}}\frac{4\pi}{\lambda}-\frac{b}{d}\right]Acos\left(\omega t\right)\\
\\
 & - & \frac{1}{2}\left[\frac{1}{\sqrt{2}}\left(\frac{4\pi}{\lambda}\right)^{2}+\frac{b}{d^{2}}\right]ABcos\left[\left(\omega+\omega'\right)t\right]\\
\\
 & - & \left[\frac{1}{\sqrt{2}}\left(\frac{1}{2}\frac{4\pi}{\lambda}\right)^{2}+\frac{1}{4}\frac{b}{d^{2}}\right]A^{2}cos\left(2\omega t\right)\\
\\
 & + & \left[\frac{1}{\sqrt{2}}\left(\frac{1}{2}\frac{4\pi}{\lambda}\right)^{4}+\frac{1}{32}\frac{b}{d^{4}}\right]A^{2}B^{2}cos\left[\left(2\omega+2\omega'\right)t\right]\\
\\
 & + & \left[\frac{1}{\sqrt{2}}\frac{4}{3}\left(\frac{4\pi}{\lambda}\right)^{4}+\frac{1}{192}\frac{b}{d^{4}}\right]A^{4}cos\left(4\omega t\right)
\end{eqnarray*}

If we define an optical contrast factor $R_{n}$ as the ratio between the scattering term in
each coefficient that goes as $1/d^{n}$ and the background artifact that goes as $1/\lambda^{n}$ of each $n$ harmonic demodulated
signal, the advantage in demodulating the signal
at certain frequencies in comparison to others becomes clear. If for example, we
calculate the ratio between $R_{2(\omega+\omega')}$ and $R_{2\omega}$
for a wavelength of 1580nm, we achieve a 160 fold contrast enhancement,
which is the same enhancement achieved in mono-modal excitation while
demodulating on the fourth harmonic, as shown in Figure \ref{fig:Figure1}(c). Thus, the same near field to background optical contrast is predicted to occur for frequencies such as:
$R_{2\omega}=R_{\omega+\omega'}$,
$R_{3\omega}=R_{2\omega+\omega'}=R_{\omega+2\omega'}$ and
$R_{4\omega}=R_{2\omega+2\omega'}$. Namely, one could achieve high optical contrast for lower demodulation frequencies, thus with a stronger signal.

An additional advantage in multifrequency excitation SNOM is the wider range of tip oscillation amplitudes suitable for eliminating far-field background from the near-field signal. This could be achieved by expanding the finding \cite{gucciardi2006far} that the total intensity of the signal measured by the
detector in a mono-modal SNOM setup produces
a non-vanishing background term at all $n$ harmonics of the signal,
which is directly proportional to $J_{n}(2ka_{1})$. In this term,
$a_{1}$ is the single tip oscillation amplitude, and $k$ is the wave vector
of the illuminating field. In order to suppress the background, one
must choose $a_{1}$ in a way that mathematically cancels this term.

Generalizing this analytic derivation to the bi-modal excitation technique results in an extension of the available solutions for background suppression from a single tip oscillation amplitude to a two-dimensional plane of possible sets of the two oscillation amplitudes for each mode of excitation. This is derived from the fact that the new background term is proportional to the \emph{product} of two Bessel
functions:
\begin{eqnarray*}
BKG_{n} & \propto & J_{n}\left(2ka_{1}\right)\times J_{n}\left(2ka_{2}\right)\\
\end{eqnarray*}
Thus, in this case, the solution space expands, and one is free to choose from a set of available tip oscillation
amplitudes $a_{1}$ and $a_{2}$, in order to completely cancel this term. It should be stressed once again that in this case, the signal should be demodulated at the composite harmonics of the two mechanical frequencies of the tip. 

\section{Experimental Results and Discussion}
\begin{figure}
\begin{minipage}{\columnwidth}
\includegraphics[width=3.0in]{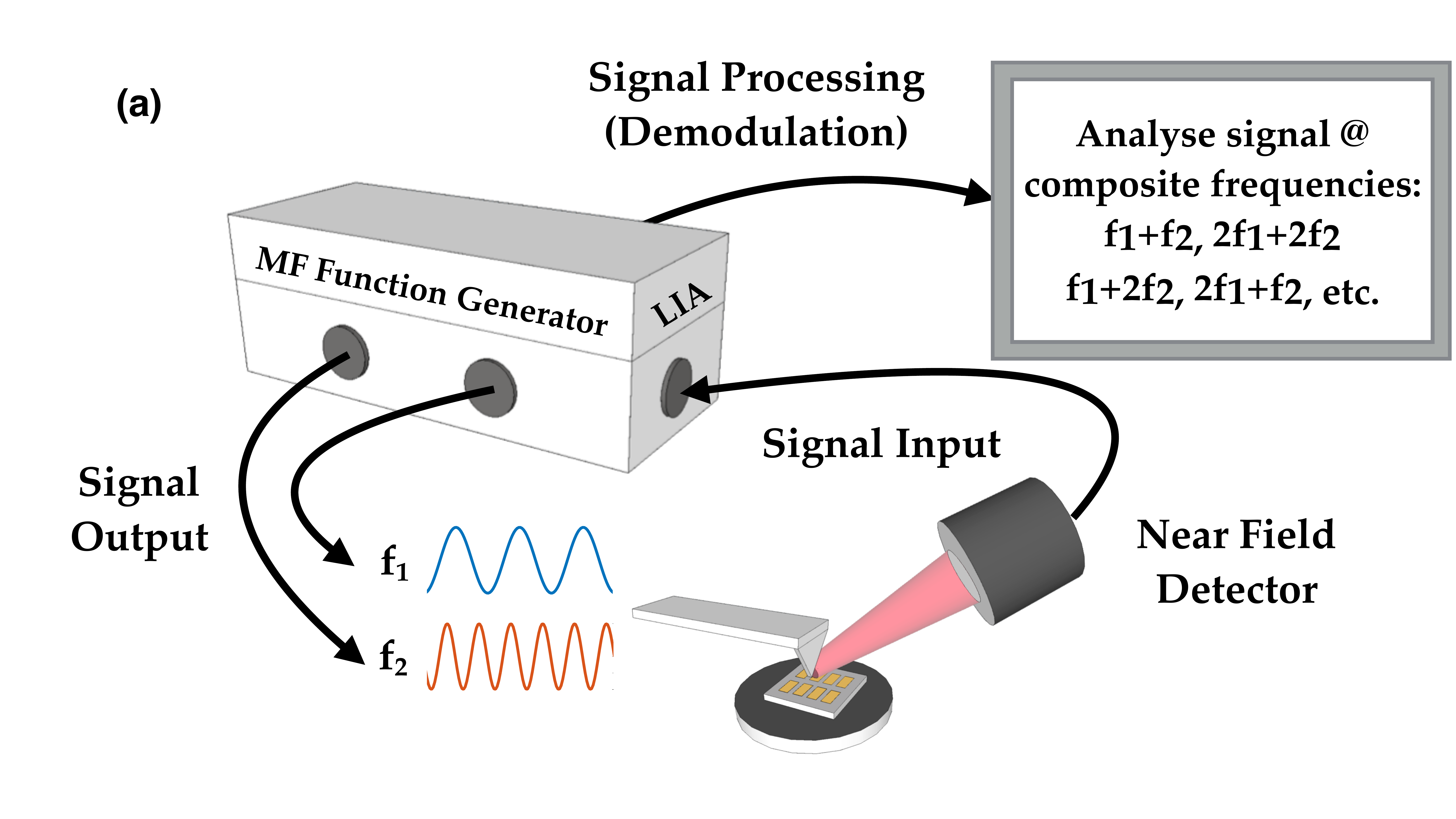}
\end{minipage}
\begin{minipage}{\columnwidth}
\includegraphics[width=3.0in]{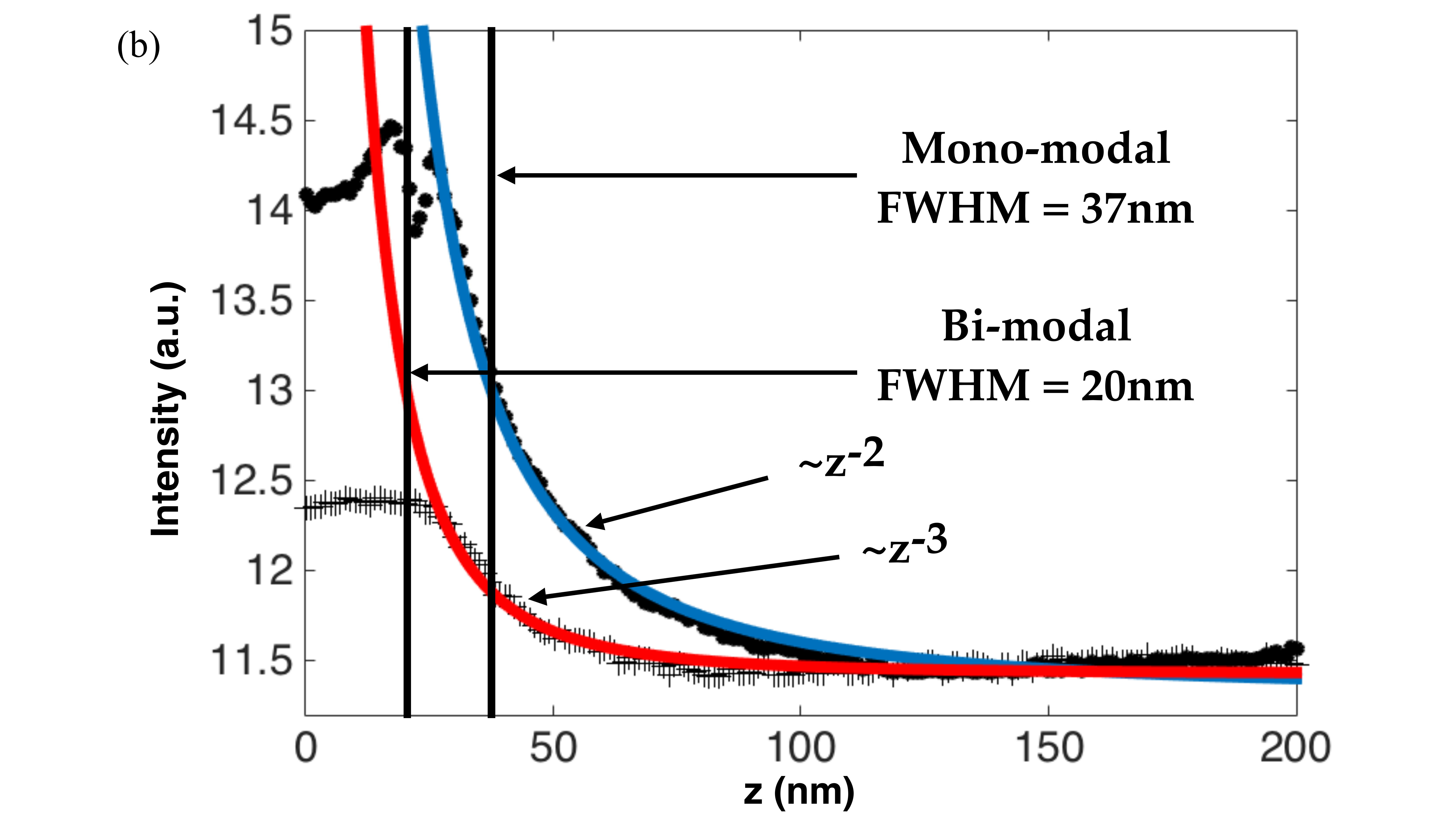}\centering
\end{minipage}
\caption{(\emph{Color online}) Experimental set-up for multifrequency SNOM measurements. (a) Scheme of experimental set-up. Zurich Instruments UHF-LI acts as function generator, to feed AFM tip with frequencies $f_{1}$ and $f_{2}$. Near field signal is detected via a multifrequency lock-in detection scheme in the Lock-In Amplifier (LIA) and demodulated at composite frequencies. (b) Near field sensitivity measurements for mono-modal SNOM vs. bi-modal SNOM. Circles: Near field signal of Au split ring resonator on ITO as a function of tip-sample distance, measured by the mono-modal technique, detected at the 4th harmonic of $f_{1}$. Crosses: Same as above, measured by the bi-modal technique, detected at the second harmonic of the sum $f_{1}+f_{2}$. Blue: Fit of mono-modal measurements to $~1/z^{2}$. Red: Fit of bi-modal measurements exhibits enhanced sensitivity and a faster rise of approach $~1/z^{3}$. The FWHM of the bi-modal rise of approach is 20nm, while it is 37nm for the mono-modal measurement.}
\label{fig:Figure2}
\end{figure}

In order to examine optical near-field measurements with the multifrequency SNOM technique, we used plasmonic nanostructure arrays, comprised of Au nanobars and split ring
resonators (SRR). These were fabricated via standard electron beam lithography, and deposited with a height of 100nm on an ITO substrate. The near-field measurements were done using a NeaSpec
neaSNOM, illuminated with a tunable CW laser (Toptica CTL1550) between 1550-1580nm. We used a Zurich Instruments UHF Lock-in Amplifier, with its many available oscillators, to externally drive the AFM cantilever on the one hand, and to demodulate the detected scattered signal at any frequency of our choice on the other hand. Figure \ref{fig:Figure2}(a) depicts a schematic representation of the experimental set-up. 

To compare the traditional mono-modal SNOM technique with our bi-modal method, we performed two different sets of measurements. First, the tip was made to mechanically vibrate mono-modally at its first flexural frequency $f_{1} = 70kHz$, a characteristic value slightly different for each tip, and the near field optical signal was collected at this frequency and its higher harmonics. Next, the bi-modal excitation method was employed; the tip was vibrated simultaneously at  \emph{two} of its first flexural frequencies, $f_{1}$ and $f_{2} = 420kHz$, and the collected near field optical signal was demodulated at multifrequency harmonics, theoretically predicted to display the same near field to background contrast $R_{n}$ as their mono-modal counterparts (see Figure \ref{fig:Figure1}(c)). Examples of these include the sum of the frequencies $f_{1}+f_{2}$, predicted to produce the same signal as $2f_{1}$, its second harmonic, equivalent to $4f_{1}$, and the sum of the higher harmonics of the two excitation frequencies, such as $f_{1}+2f_{2}$, equivalent to $3f_{1}$. 

In order to test the near field sensitivity of the technique,
we initially measured the detected near field
signal of a single point of high signal intensity in each nanostructure as a function of the tip-sample
distance. Figure \ref{fig:Figure2}(b) is a comparison of these measurements, employing
the standard mono-modal method, demodulated at the fourth
harmonic of the tip's first flexural frequency $4f_{1}$ vs. the bi-modal technique, demodulated at the second harmonic of the
sum of the tip's first and second flexural frequencies $2(f_{1}+f_{2})$.
It is apparent that the bi-modal method exhibits the desirable tip
sharpening effect \cite{knoll2000enhanced}, as its narrower signal abruptly changes closer to the sample, with a
faster rise of approach of $1/z^{3}$, compared to $1/z^{2}$, fit
to the mono-modal measurements. A further support for this claim is the twofold enhancement in the sensitivity of this measurement exhibited by the decrease in the FWHM. The FWHM marked in the bi-modal measurement is $20nm$, roughly half of that extracted from the mono-modal measurement, $37nm$. Namely, this implies that the signal detected via the mono-modal method at $z=35nm$ (blue curve in \ref{fig:Figure2}(b)), for example, contains optical background, that is absent from the signal detected via the bi-modal method (red curve in \ref{fig:Figure2}(b)).

\begin{figure}
\includegraphics[scale=0.13]{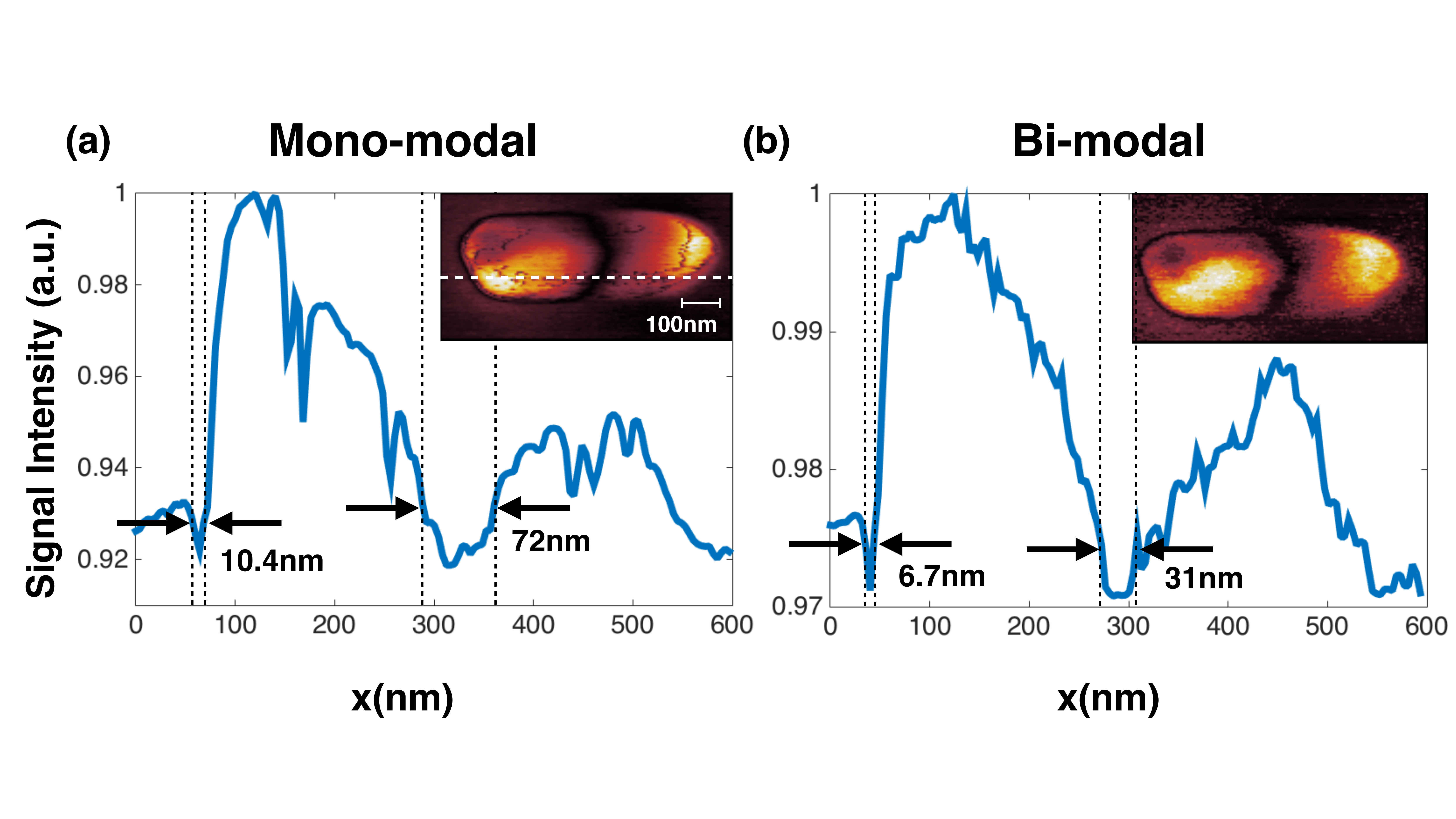}
\protect\caption{(Color online) Comparison of near-field optical image of a gold nanobar fabricated on ITO, obtained by (a) mono-modal excitation and detection on $4f_{1}$ and (b) bi-modal excitation and detection on $2f_{1}+2f_{2}$. The line plots of the signal intensity as a function of lateral distance, measured along the white dotted line in the inset images. The marked FWHM show a decrease in the bi-modal measurement, implying an improvement in spatial resolution.   \label{fig:Figure3}}
\end{figure} 

To examine whether the enhanced sensitivity translates into improved resolution, we performed a complete near-field scan of a 540nm Au nanobar, illuminated with
polarization along its long axis, employing both techniques, shown in \ref{fig:Figure3}. The inset of Figure \ref{fig:Figure3}(a) is a near field image obtained by mono-modal excitation and signal demodulation at $4f_{1}$. The inset of Figure \ref{fig:Figure3}(b) is the same image obtained by bi-modal excitation and signal demodulation at $2f_{1}+2f_{2}$, which is the frequency value theoretically predicted to produce the same near field optical contrast (see Figure \ref{fig:Figure1}(c)). Note two low intensity points, located at the left edge and the middle of the sample, that appear sharper in the image obtained by the bi-modal excitation method. We quantified
this by plotting the signal intensity, proportional to the near field
scattering, as a function of lateral distance, measured along the white dotted line in each of the above images.

The low intensity points in each figure are depicted by dips in these
line plots. While the FWHM of the left dip obtained in Figure \ref{fig:Figure3}(a) is
10.4nm, it is narrowed down to 6.7nm in the bi-modal measurement presented in Figure \ref{fig:Figure3}(b). Moreover, the FWHM of the middle dip decreases by a factor of over 2 in the bi-modal scheme, implying
an increase in spatial resolution. This finding complies with the notion that the spatial resolution of SNOM measurements is directly proportional to the $z$ axis sensitivity \cite{1991SurSc.253..353K}. The increase in resolution is consistent, although not as prominent, for for third harmonic measurements, obtained via mono-modal excitation and detection at $3f_{1}$ and  via bi-modal excitation and detection at $f_{1}+2f_{2}$. 

Furthermore, in the above comparisons, the mono-modal excitation measurements exhibit so-called "$z$ artifacts". These are parasitical low intensity streaks in the optical signal, originating from topographical features of the nanobars, that do not appear in the bi-modal measurements. 

\begin{figure}
\includegraphics[width = 3.0in]{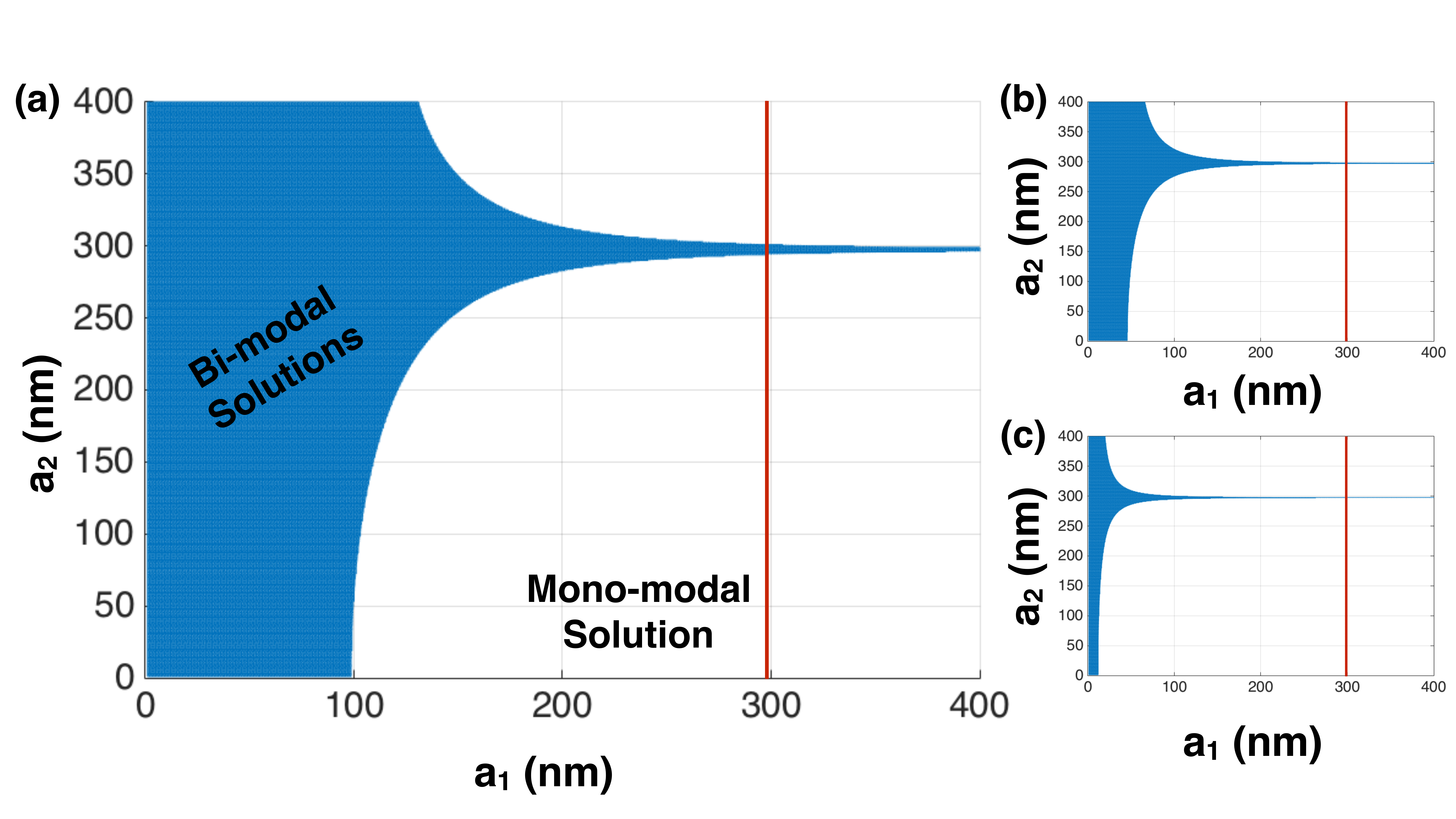}
\protect\caption{(Color online) Numerical simulation of tip oscillation amplitude values for complete optical background signal suppression in near field measurements, illuminated at $\lambda=1550nm$; (a) Red: Single value of tip oscillation amplitude for background suppression $a_{1}=100nm$ in mono-modal excitation for signal demodulation at $4f_{1}$. Blue: Range of tip oscillation amplitudes for background suppression $(a_{1},a_{2})$ in bi-modal excitation and signal demodulation at $2(f_{1}+f_{2})$. (b) Red: Mono-modal excitation for signal demodulation at $3f_{1}$. Blue: Bi-modal excitation for signal demodulation at $f_{1}+2f_{2}$. (c) Red: Mono-modal excitation for signal demodulation at $2f_{1}$. Blue: Bi-modal excitation for signal demodulation at $f_{1}+f_{2}$. \label{fig:Figure4}}
\end{figure}

An additional experimental advantage to multifrequency SNOM mentioned earlier is the fact
that the optical background is proportional to $ J_{n}(2ka_{1})\times J_{n}(2ka_{2})$.
Figure \ref{fig:Figure4}(a)  is a theoretical simulation of the tip oscillation amplitude
values for complete background suppression, calculated for the experimental
variables of the images in Figure \ref{fig:Figure3}, including an illumination wavelength of $\lambda=1550nm$.  

The red line in Figure\ref{fig:Figure4}(a) represents the single
value of this amplitude in the
mono-modal case of $a_{1} \approx 300nm$. The blue range of values are the
tip oscillation amplitude pairs $(a_{1},a_{2})$ one could chose in
order to suppress the background contribution to near-field measurements
using the bi-modal technique. Here, we chose values of $\approx(100nm,100nm)$,
in order to maximize the near-field signal, while ensuring that this
amplitude is at least a factor smaller than the illuminating wavelength, so that the theoretical model holds. Figure \ref{fig:Figure4}(b) is the same calculation, where the signal is demodulated at $3f_{1}$ in the mono-modal case, and at $f_{1}+2f_{2}$ in the bi-modal case. In Figure \ref{fig:Figure4}(c) the signal is demodulated at $2f_{1}$ in the mono-modal case, and at $f_{1}+f_{2}$ in the bi-modal case. The range of available tip oscillation amplitudes expands as the bi-modal demodulation frequencies rises, where the single oscillation amplitude, which is a function of the illuminating wavelength, remains the same.

It should be noted that
this multifrequency technique is also compatible with the aforementioned pseudoheterodyne
detection scheme\cite{ocelic2006}. This compatibility is under the condition that the reference wave is not an integer product of either of the tip's excitation frequencies. 

\section{Conclusions}

We have theoretically introduced a novel multifrequency excitation and demodulation
technique to efficiently extract a near-field signal with improved
sensitivity and deep sub-wavelength resolution reaching $\lambda/230$. Our experimental results demonstrate an enhanced tip-sharpening effect for bi-modal excitation vs. mono-modal excitation leading to improved spatial resolution. This
is a comprehensive and feasible experimental method due to its many degrees of freedom, resulting in background suppression and
increased optical contrast with a high SNR. The richness of the technique allows to expand the conventional near field scattering type method to detect weaker near field signals at lower demodulation harmonics, thus enabling their thorough measurement.   Our proof of concept breaks the ground for an unmatched capability of near field optical detection, without compromising the sub-wavelength spatial resolution. 

\bibliography{mfnsom}

\end{document}